\input harvmac


 \def\quad{{\ \ }}
 
 \def\ads{AdS$_5\times$S$^5$}
 \def\op{${\cal O}_\Sigma$}

\def\lra{\Leftrightarrow}

\def\Tr{{\rm Tr}}

\def\op{${\cal O}_{\Sigma}$}
\def\ope{{\cal O}_{\Sigma}}

\let\includefigures=\iftrue
\newfam\black
\includefigures
\input epsf
\def\figin{\epsfcheck\figin}\def\figins{\epsfcheck\figins}
\def\epsfcheck{\ifx\epsfbox\UnDeFiNeD
\message{(NO epsf.tex, FIGURES WILL BE IGNORED)}
\gdef\figin##1{\vskip2in}\gdef\figins##1{\hskip.5in}
\else\message{(FIGURES WILL BE INCLUDED)}%
\gdef\figin##1{##1}\gdef\figins##1{##1}\fi}
\def\DefWarn#1{}
\def\figinsert{\goodbreak\midinsert}
\def\ifig#1#2#3{\DefWarn#1\xdef#1{fig.~\the\figno}
\writedef{#1\leftbracket fig.\noexpand~\the\figno}%
\figinsert\figin{\centerline{#3}}\medskip\centerline{\vbox{\baselineskip12pt
\advance\hsize by -1truein\noindent\footnotefont{\bf Fig.~\the\figno:}
#2}}
\bigskip\endinsert\global\advance\figno by1}
\else
\def\ifig#1#2#3{\xdef#1{fig.~\the\figno}
\writedef{#1\leftbracket fig.\noexpand~\the\figno}%
#2}}
\global\advance\figno by1}
\fi


\def\sym{  \> {\vcenter  {\vbox
                  {\hrule height.6pt
                   \hbox {\vrule width.6pt  height5pt
                          \kern5pt
                          \vrule width.6pt  height5pt
                          \kern5pt
                          \vrule width.6pt height5pt}
                   \hrule height.6pt}
                             }
                  } \>
               }
\def\fund{  \> {\vcenter  {\vbox
                  {\hrule height.6pt
                   \hbox {\vrule width.6pt  height5pt
                          \kern5pt
                          \vrule width.6pt  height5pt }
                   \hrule height.6pt}
                             }
                       } \>
               }
\def\anti{ \>  {\vcenter  {\vbox
                  {\hrule height.6pt
                   \hbox {\vrule width.6pt  height5pt
                          \kern5pt
                          \vrule width.6pt  height5pt }
                   \hrule height.6pt
                   \hbox {\vrule width.6pt  height5pt
                          \kern5pt
                          \vrule width.6pt  height5pt }
                   \hrule height.6pt}
                             }
                  } \>
               }

\Title{\vbox{\baselineskip12pt\hbox{arXiv:0704.1657}}}
{\vbox{\centerline{
Bubbling Surface Operators} \vskip6pt\centerline
{And S-Duality}}}

\vskip-5pt
\centerline{Jaume Gomis$^1$ and Shunji Matsuura$^2$}
 
\smallskip
\smallskip
\bigskip
\centerline{\it Perimeter Institute for Theoretical Physics}
\centerline{\it Waterloo, Ontario N2L 2Y5, Canada$^{1,2}$ 
}
\footnote{${}^{}$}{${}^{1}$\tt jgomis@perimeterinstitute.ca}
\footnote{${}^{}$}{${}^{2}$\tt smatsuura@perimeterinstitute.ca}
\medskip
\centerline{\it Department of Physics}
\centerline{\it University of Tokyo,  7-3-1 Hongo, Tokyo$^2$}
\vskip .3in
\centerline{\bf{Abstract}}
\medskip
\medskip
We construct  smooth asymptotically \ads\ solutions of Type IIB supergravity corresponding  to all the half-BPS surface operators in ${\cal N}=4$ SYM. All the parameters labeling a half-BPS surface operator  are identified in the corresponding bubbling geometry. We use the supergravity description of surface operators   to study the action of the $SL(2,Z)$ duality group of ${\cal N}=4$ SYM on the parameters of the surface operator, and find that it coincides with the recent proposal by Gukov and Witten in the framework of the gauge theory approach to the  geometrical Langlands  with ramification. We also show that whenever
a bubbling geometry  becomes singular  that the path integral description of the corresponding 
surface operator  also becomes singular.

\Date{04/2007}

\newsec{Introduction and Summary}

Gauge invariant operators play a central role in  the   gauge theory holographically describing  quantum gravity with     AdS boundary conditions
\lref\MaldacenaRE{
  J.~M.~Maldacena,
  ``The large N limit of superconformal field theories and supergravity,''
  Adv.\ Theor.\ Math.\ Phys.\  {\bf 2}, 231 (1998)
  [Int.\ J.\ Theor.\ Phys.\  {\bf 38}, 1113 (1999)]
  [arXiv:hep-th/9711200].
}
\lref\GubserBC{
  S.~S.~Gubser, I.~R.~Klebanov and A.~M.~Polyakov,
  ``Gauge theory correlators from non-critical string theory,''
  Phys.\ Lett.\  B {\bf 428}, 105 (1998)
  [arXiv:hep-th/9802109].
} 
\lref\WittenQJ{
  E.~Witten,
  ``Anti-de Sitter space and holography,''
  Adv.\ Theor.\ Math.\ Phys.\  {\bf 2}, 253 (1998)
  [arXiv:hep-th/9802150].
}
\MaldacenaRE\GubserBC\WittenQJ , as   correlation functions
of gauge invariant operators are the only   observables in the  boundary gauge theory. Finding the bulk description of all gauge invariant operators is   necessary in order to be able to formulate an arbitrary  bulk experiment   in terms of  gauge theory variables.  

In this paper we  provide the bulk description of a novel class  of half-BPS operators in ${\cal N}=4$ SYM
\lref\GukovJK{
  S.~Gukov and E.~Witten,
  ``Gauge theory, ramification, and the geometric langlands program,''
  arXiv:hep-th/0612073.
}
which are supported on a surface $\Sigma$ \GukovJK. 
These nonlocal surface operators
\op\  are defined by quantizing ${\cal N}=4$ SYM in the presence of a certain codimension two singularity for the classical fields of ${\cal N}=4$ SYM. The singularity  characterizing such a   surface operator \op\  depends on $4M$ real parameters, where $M$ is the number of $U(1)$'s left unbroken by \op. Surface operators  are a higher dimensional generalization of Wilson and 't Hooft operators, which are supported on curves   and induce a codimension three singularity for the classical fields appearing in the Lagrangian. In this paper we extend  the bulk description 
of all half-BPS Wilson loop operators found  in 
\lref\GomisSB{
  J.~Gomis and F.~Passerini,
  ``Holographic Wilson loops,''
  JHEP {\bf 0608}, 074 (2006)
  [arXiv:hep-th/0604007].
}
\GomisSB\ (see also\foot{The description of Wilson loops in the fundamental representation goes back to 
\lref\MaldacenaIM{
  J.~M.~Maldacena,
   ``Wilson loops in large N field theories,''
  Phys.\ Rev.\ Lett.\  {\bf 80}, 4859 (1998)
  [arXiv:hep-th/9803002].
}
\lref\ReyIK{
  S.~J.~Rey and J.~T.~Yee,
   ``Macroscopic strings as heavy quarks in large N gauge theory and  anti-de
   Sitter supergravity,''
  Eur.\ Phys.\ J.\  C {\bf 22}, 379 (2001)
  [arXiv:hep-th/9803001].
}
\ReyIK\MaldacenaIM.}
\lref\DrukkerKX{
  N.~Drukker and B.~Fiol,
  JHEP {\bf 0502}, 010 (2005)
  [arXiv:hep-th/0501109].
}
\lref\YamaguchiTE{
  S.~Yamaguchi,
  ``Bubbling geometries for half BPS Wilson lines,''
\hskip-3pt  arXiv:hep-th/0601089.
}
\lref\YamaguchiTQ{
  S.~Yamaguchi,
  ``Wilson loops of anti-symmetric representation and D5-branes,''
  JHEP {\bf 0605}, 037 (2006)
  [arXiv:hep-th/0603208].
}
\lref\LuninXR{
  O.~Lunin,
  ``On gravitational description of Wilson lines,''
  JHEP {\bf 0606}, 026 (2006)
  [arXiv:hep-th/0604133].
}
\lref\GomisIM{
  J.~Gomis and F.~Passerini,
  ``Wilson loops as D3-branes,''
  JHEP {\bf 0701}, 097 (2007)
  [arXiv:hep-th/0612022].
}
\DrukkerKX\YamaguchiTE\YamaguchiTQ\LuninXR\GomisIM)
to all half-BPS surface operators.


We find the  asymptotically \ads\ solutions  of Type IIB supergravity  corresponding to   all  half-BPS surface operators  \op\ in ${\cal N}=4$ $U(N)$ SYM. The topology and geometry of the ``bubbling" solution  is  completely determined in terms of some   data, very much like in  the case studied by  Lin, Lunin and Maldacena (LLM) in the context of half-BPS local operators
\lref\LinNB{
  H.~Lin, O.~Lunin and J.~M.~Maldacena,
  ``Bubbling AdS space and 1/2 BPS geometries,''
  JHEP {\bf 0410}, 025 (2004)
  [arXiv:hep-th/0409174].
}
\LinNB\foot{The bubbling geometry description of half-BPS Wilson loops 
was found in \YamaguchiTQ\LuninXR\ while that of half-BPS
domain wall operators was found in 
\lref\GomisCU{
  J.~Gomis and C.~Romelsberger,
   ``Bubbling defect CFT's,''
  JHEP {\bf 0608}, 050 (2006)
  [arXiv:hep-th/0604155].
}
\LuninXR\GomisCU. For the bubbling Calabi-Yau geometries for Wilson loops in Chern-Simons, see
\lref\GomisMV{
  J.~Gomis and T.~Okuda,
  ``Wilson loops, geometric transitions and bubbling Calabi-Yau's,''
  JHEP {\bf 0702}, 083 (2007)
  [arXiv:hep-th/0612190].
}
 \GomisMV.}. 
In fact, we identify the  system of equations determining the supergravity solution 
 corresponding to  the half-BPS surface operators in ${\cal N}=4$ SYM with that obtained by   ``analytic" continuation of the LLM equations
\lref\LinNH{
  H.~Lin and J.~M.~Maldacena,
  ``Fivebranes from gauge theory,''
  Phys.\ Rev.\  D {\bf 74}, 084014 (2006)
  [arXiv:hep-th/0509235].
}
\LinNB \LinNH.

The data determining the topology and geometry of a supergravity solution is characterized  by  the   position  of a collection of $M$ point 
particles 
 in a three dimensional space $X$, where $X$ is a submanifold of  the 
ten dimensional geometry. Different particle configurations give rise to different asymptotically \ads\ geometries.
\medskip\medskip\medskip\medskip
\ifig\Youngtabcolumnsrows{$a)$ The metric and five-form flux is determined once the position of the particles  in $X$ 
-- labeled by coordinates $(\vec{x}_l,y_l)$ where $y\geq 0$  --  is given. The $l$-th particle is associated with a point $P_l\in X$.  $b$) The configuration corresponding to the \ads\ vacuum.}
{\epsfxsize3in\epsfbox{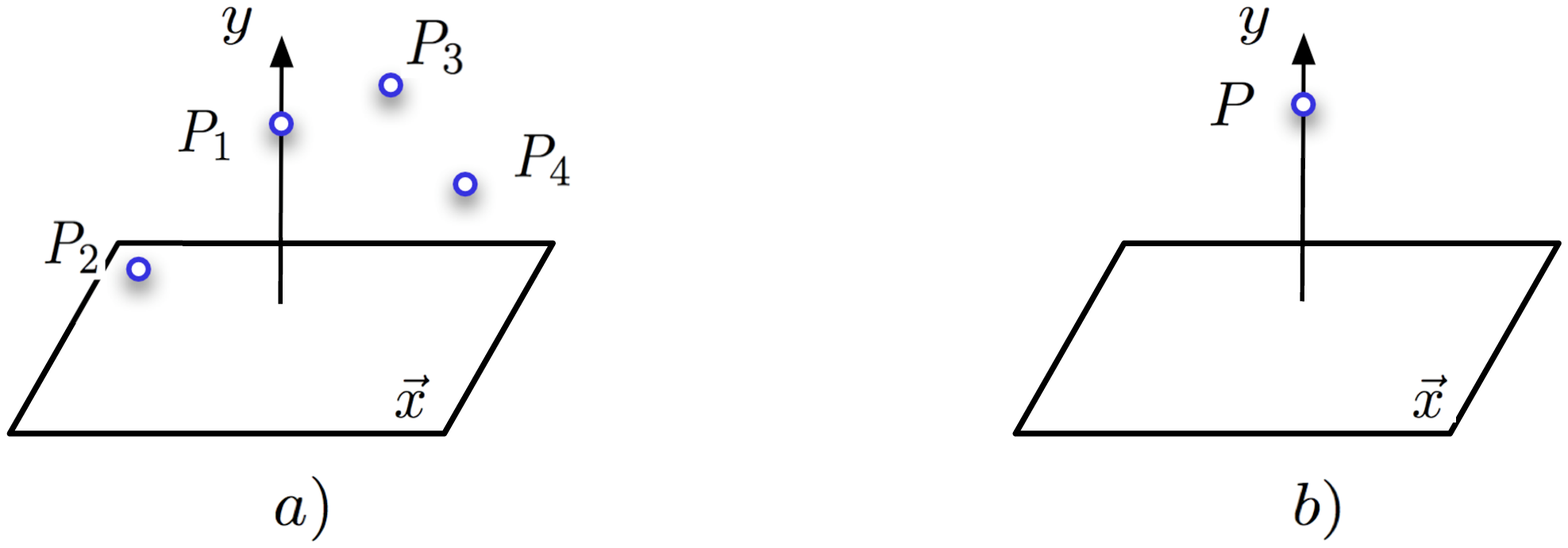}}

Even though the choice of a particle distribution in $X$ completely determines the geometry  and topology of the metric   and the corresponding RR five-form field strength, further choices have to be made to fully characterize a solution  of Type IIB supergravity on this geometry\foot{This is on top of the obvious choice of  dilaton and axion, which gets identified with   the complexified coupling constant in ${\cal N}=4$ SYM.}.

Given a configuration of $M$ particles in $X$, the corresponding ten dimensional geometry develops
$M$ non-trivial disks which end on the boundary\foot{The conformal boundary in this case is AdS$_3\times$S$^1$, where surface operators in ${\cal N}=4$ SYM can be studied  by specifying non-trivial boundary conditions.} of \ads\ on a non-contractible S$^1$. Since Type IIB supergravity has two two-form gauge fields, one from the NS-NS sector and one from the RR sector, a  solution of the Type IIB supergravity equations of motion is fully determined only once the holonomy of the  two-forms    around the various  disks  is specified:
\eqn\holo{
 \int_{D_l} {B_{NS}\over 2\pi}\qquad \qquad  \int_{D_l} {B_{R}\over 2\pi}\qquad l=1,\ldots,M.}
 Therefore, an asymptotically \ads\ solution depends on the position of the $M$ particles  in $X$  -- given by $(\vec{x}_l,y_l)$ --
  and on the holonomies of the two-forms \holo. 

A precise dictionary is given between all  the $4M$ parameters that label a half-BPS surface operator \op\ and all the parameters describing the corresponding   supergravity solution. We    show that the supergravity solution   describing a    half-BPS surface operator is regular and that whenever the  supergravity solution develops a singularity  
 the ${\cal N}=4$ SYM path integral description of the corresponding surface operator also develops a singularity.



We study the action of the $SL(2,Z)$ symmetry of Type IIB string theory on the supergravity solutions representing the half-BPS surface operators in ${\cal N}=4$ SYM. By using the proposed dictionary between the parameters of a supergravity solution and the parameters of the corresponding  surface operator, we can show that the action of  $S$-duality induced on the parameters of a surface operator coincides with the recent proposal by Gukov and Witten \GukovJK\ in the framework of the gauge theory approach to the  geometrical Langlands 
\lref\KapustinPK{
  A.~Kapustin and E.~Witten,
  ``Electric-magnetic duality and the geometric Langlands program,''
 arXiv:hep-th/0604151.
  }
\KapustinPK\foot{See e.g. 
\lref\FrenkelPA{
  E.~Frenkel,
  ``Lectures on the Langlands program and conformal field theory,''
  arXiv:hep-th/0512172.
}
\FrenkelPA\ for a review of the geometric Langlands program.}  with ramification.

 Whether  surface operators can serve as novel order parameters in gauge theory remains an important open question. It is our hope that the viewpoint on these operators provided by the supergravity solutions in this paper may help shed  light on this crucial question.

The plan of the rest of the paper is as follows. In section $2$ we study the gauge theory singularities corresponding to surface operators in ${\cal N}=4$ SYM, study the symmetries preserved by a half-BPS surface operator and review the  proposal in \GukovJK\ for the action of $S$-duality on the parameters that a half-BPS surface operator depends on. We also compute the scaling weight of these operators and show that it is invariant under Montonen-Olive duality. In section $3$ we construct the solutions of Type IIB supergravity describing the half-BPS surface operators. We identify all the parameters that a surface operator depends on in the supergravity solution and show that the action of $S$-duality on surface operators proposed in \GukovJK\ follows from the action of $SL(2,Z)$ on the classical solutions of supergravity. The  Appendices contain some   details omitted in the main text.




\newsec{Surface Operators in Gauge Theories}

 A surface operator ${\cal O}_\Sigma$ is labeled by a surface 
 $\Sigma$ in $R^{1,3}$  and by a conjugacy class $U$ of the gauge group $G$.  The data that characterizes a surface operator 
 ${\cal O}_\Sigma$, the surface $\Sigma$ and the conjugacy class $U$, can be identified with that of an external  string used to probe the theory. The surface $\Sigma$ corresponds to the worldsheet of a string while the conjugacy class  $U$ is associated  to the Aharonov-Bohm phase acquired by a charged particle encircling the string.

 The singularity\foot{Previous work involving codimension two singularities in gauge theory include
\lref\RohmXK{
  R.~M.~Rohm,
  ``Some Current Problems In Particle Physics Beyond The Standard Model,''
}
\lref\PreskillBM{
  J.~Preskill and L.~M.~Krauss,
  ``Local Discrete Symmetry And Quantum Mechanical Hair,''
  Nucl.\ Phys.\  B {\bf 341}, 50 (1990).
}
\lref\AlfordYX{
  M.~G.~Alford, K.~M.~Lee, J.~March-Russell and J.~Preskill,
  Nucl.\ Phys.\  B {\bf 384}, 251 (1992)
  [arXiv:hep-th/9112038].
}
\lref\GomisCE{
  J.~Gomis,
   ``Anomaly cancellation in noncritical string theory,''
  JHEP {\bf 0510}, 095 (2005)
  [arXiv:hep-th/0508132].
}
\RohmXK\PreskillBM\AlfordYX.} 
 in the gauge field produced by a surface operator
is that of a non-abelian vortex. This singularity in the gauge field
can be characterized by the phase acquired by a charged particle circumnavigating around the string. This gives rise to a group element\foot{We now focus on $G=U(N)$ as it is the relevant gauge group for describing string theory with asymptotically  AdS$_5\times$S$^5$ boundary conditions.}
 $U\subset  U(N)$ 
\eqn\holoa{
U\equiv P \exp {i\oint A}\subset  U(N),}
which corresponds to the Aharonov-Bohm phase picked up by the wavefunction of the charged particle. Since gauge transformations act by conjugation $U\rightarrow gUg^{-1}$, a surface operator is labeled by a conjugacy class of the gauge group.

By performing a gauge transformation, the matrix $U$ can be   diagonalized. If we demand that the gauge field configuration is scale invariant -- so that \op\ has a well defined scaling weight --  
then the   gauge field produced by a surface operator can then be written as 
\smallskip
\eqn\holo{
A= \pmatrix{\alpha_{1}\otimes 1_{N_1}&0 &\ldots & 0\cr
0& \alpha_{2}\otimes1_{N_2}&\ldots & 0 \cr
\vdots &\vdots &\ddots&\vdots\cr
0&0&\ldots&\alpha_{M}\otimes1_{N_M}}d\theta,}
\smallskip\smallskip
\smallskip
\noindent
where $\theta$ is the polar angle in the $R^2\subset R^{1,3}$ plane normal to $\Sigma$  and  $1_n$ is the $n$-dimensional unit matrix. 
We note that the matrix $U$ takes values on the maximal torus $T^N=R^N/Z^N$ of the $U(N)$ gauge group. Therefore the parameters $\alpha_i$  take values on a circle of unit radius. 

The surface operator corresponding to \holo\ spontaneously breaks the $U(N)$ gauge symmetry   along $\Sigma$ down to the so called Levi  group $L$, where a group of Levi type is  characterized by 
the subgroup of $U(N)$ that commutes with \holo. Therefore, $L = \prod_{l=1}^M U(N_l)$, where  $N=\sum_{l=1}^M N_l$.

Since the gauge group is broken down to the Levi group $L = \prod_{l=1}^M U(N_l)$ along $\Sigma$, there is a further choice \GukovJK\ in the definition  of   \op\ consistent with the symmetries and equations of motion.  This corresponds to turning on a two dimensional $\theta$-angle for the 
unbroken $U(1)$'s along the string worldsheet $\Sigma$. The associated operator insertion into the ${\cal N}=4$ SYM path integral is    given by:
\eqn\insertion{
\exp\left(i\sum_{l=1}^M\eta_l\int_\Sigma\hbox{Tr}\;{F_l}\right).}

 The parameters  $\eta_i$ takes values in the maximal torus of the $S$-dual or Langlands dual gauge group $^{L}G$ \GukovJK. Therefore, since $^{L}G=U(N)$ for $G=U(N)$,   we have that the matrix of $\theta$-angles of a 
 surface operator  \op\  characterized by the Levi group $L = \prod_{l=1}^M U(N_l)$ is given by the $L$-invariant matrix:
\smallskip
\eqn\angle{
\eta= \pmatrix{{\eta_{1}}\otimes 1_{N_1}&0 &\ldots & 0\cr
0& {\eta_{2}}\otimes1_{N_2}&\ldots & 0 \cr
\vdots &\vdots &\ddots&\vdots\cr
0&0&\ldots&{\eta_{M}}\otimes1_{N_M}}.}
\smallskip
\smallskip
\noindent
The parameters $\eta_i$, being two dimensional $\theta$-angles, 
also take values on a circle of unit radius.



Therefore, a surface operator \op\ in pure gauge theory with Levi group $L = \prod_{l=1}^M U(N_l)$ is labeled by $2M$ $L$-invariant parameters $(\alpha_l,\eta_l)$ up to the action of $S_M$, which acts by permuting the different eigenvalues  in   \holo\ and \angle.
The operator is then  defined by expanding the path integral with the  insertion of  the operator \insertion\ around the singularity \holo, 
and by  integrating over connections that are smooth near $\Sigma$. In performing the path integral, we must divide \GukovJK\ by the gauge transformations that take values in $L =\prod_{l=1}^M U(N_l)$ when restricted to $\Sigma$. This means that the operator becomes singular whenever  the unbroken gauge symmetry near $\Sigma$ gets enhanced, corresponding to when eigenvalues in \holo\ and \angle\ coincide.

\medskip
\medskip
\noindent
{\it Surface Operators in ${\cal N}=4$ SYM}
 \medskip
\medskip

In a gauge theory with extra classical  fields like ${\cal N}=4$ SYM, the surface operator ${\cal O}_\Sigma$ may    produce  a singularity for  the extra fields near the location of the surface operator. The only requirement is that 
 the singular field configuration  
solves the equations of motion of the theory away\foot{For  pure gauge theory, the field configuration in \holo\ does satisfy the Yang-Mills equation of motion $D_m F^{mn}=0$ away from $\Sigma$. Moreover, adding the two  dimensional $\theta$-angles \insertion\ does not change the equations of motion.}
 from the surface $\Sigma$. The global symmetries imposed on the operator 
\op\ determine which classical  fields in the Lagrangian develop a singularity near $\Sigma$ together with the type of singularity.

A complementary viewpoint on surface operators is to add new degrees of freedom on the surface $\Sigma$. Such an approach to surface operators in ${\cal N}=4$ SYM has been considered in
\lref\KapustinPB{
  A.~Kapustin and S.~Sethi,
  ``The Higgs branch of impurity theories,''
  Adv.\ Theor.\ Math.\ Phys.\  {\bf 2}, 571 (1998)
  [arXiv:hep-th/9804027].
}
\lref\ConstableXT{
  N.~R.~Constable, J.~Erdmenger, Z.~Guralnik and I.~Kirsch,
  ``Intersecting D3-branes and holography,''
  Phys.\ Rev.\  D {\bf 68}, 106007 (2003)
  [arXiv:hep-th/0211222].
}
\KapustinPB\ConstableXT\ where the new degrees of freedom arise from localized open strings on a brane intersection.

The basic effect of \op\  is to generate an Aharonov-Bohm phase corresponding to a group element $U$ \holoa.  If we let $z$  be the complex coordinate  in the $R^2\subset R^{1,3}$  plane normal to $\Sigma$, the singularity in the  gauge field 
configuration is then given by 
\eqn\gaugefield{
A_z=\sum_{I=1}^k{A_I\over z^I},}
where $A_I$ are   constant matrices. Scale invariance of the singularity -- which we are going to impose -- restricts $A_I=0$ for $I\geq 2$.

The operator \op\   can also excite a complex scalar field $\Phi$ of ${\cal N}=4$ SYM near $\Sigma$ while preserving half of the Poincare supersymmetries of ${\cal N}=4$ SYM.  Imposing that the singularity is scale invariant\foot{If we relax the restriction of scale invariance, one can then get other supersymmetric singularities with higher order poles   $\Phi=\sum_{I=1}^k{\Phi_I\over z^I}$ and $A$ \gaugefield. The surface operators associated with these singularities may be relevant \GukovJK\ for the  gauge theory approach to the study of the 
geometric  Langlands program with wild ramification.} 
yields
\eqn\scaleinv{
\Phi={\Phi_1\over z},}
where $\Phi_1$ is a constant matrix.


 A surface operator ${\cal O}_\Sigma$ is characterized by the  choice of an unbroken gauge group $L\subset G$  along $\Sigma$. Correspondingly, 
  the singularity of all the fields excited by \op\ must  be invariant under the unbroken gauge group $L$. 
For $L =\prod_{l=1}^M U(N_l)\subset  U(N)$   the singularity in the gauge field is the non-abelian vortex configuration in \holo\ and the two dimensional $\theta$-angles are given by \angle. $L$-invariance together with scale invariance 
 requires that $\Phi$ develops an $L$-invariant pole near $\Sigma$:   
 \smallskip
\eqn\pole{
\Phi={1\over \sqrt{2}z}\pmatrix{\beta_{1}+i\gamma_1\otimes 1_{N_1}&0 &\ldots & 0\cr
0& \beta_{2}+i\gamma_2\otimes1_{N_2}&\ldots & 0 \cr
\vdots &\vdots &\ddots&\vdots\cr
0&0&\ldots&\beta_{M}+i\gamma_M\otimes1_{N_M}}.}
\smallskip
\smallskip

Therefore, a half-BPS surface operator \op\ in ${\cal N}=4$ SYM
with Levi group $L=\prod_{l=1}^M U(N_l)$ is labeled by $4M$ $L$-invariant parameters $(\alpha_l,\beta_l,\gamma_l,\eta_l)$ up to the action of $S_M$, which permutes the different eigenvalues in \holo\angle\pole. 
The operator is    defined by the  path integral of ${\cal N}=4$ SYM
with the insertion of the operator \insertion\   expanded around the $L$-invariant singularities \holo\pole\ and by integrating over smooth fields near $\Sigma$. As in the pure gauge theory case, we must mode out by gauge transformations that take values in $L\subset  U(N)$ when restricted to $\Sigma$. The surface operator \op\ becomes singular whenever the the parameters that label the surface operator $(\alpha_l,\beta_l,\gamma_l,\eta_l)$ for $l=1,\ldots, M$ are such that they are invariant under a larger symmetry than $L$, the group of gauge transformations we have to mode out when evaluating the path integral.

\medskip
\medskip
\noindent
{\it $S$-duality of Surface Operators}
\medskip
\medskip

In ${\cal N}=4$ SYM the coupling constant combines with the four dimensional $\theta$-angle  into a complex parameter taking values in the upper half-plane:
\eqn\coupling{
\tau={\theta\over 2\pi}+{4\pi i\over g^2}.}
The group of duality symmetries of ${\cal N}=4$ SYM is an infinite discrete subgroup of $SL(2,R)$, which depends on the gauge group $G$. For ${\cal N}=4$  SYM with $G=U(N)$   the relevant symmetry group is  $SL(2,Z)$:
\eqn\trans{
{\cal M}=\pmatrix{a&b\cr
c&d}\in SL(2,Z).} 
Under $S$-duality $\tau\rightarrow -1/\tau$ and $G$ gets mapped\foot{For $G$ not a simply-laced group,
$\tau\rightarrow -1/n\tau$, where $n$ is the ratio of the length-squared of the long  and short roots of $G$.} to the $S$-dual or  Langlands dual gauge group $^{L}G$. For $G=U(N)$ the $S$-dual group is $^{L}G=U(N)$, and $SL(2,Z)$ is a symmetry of the  theory, which acts on the coupling of the theory
  by fractional linear transformations:
\eqn\fracton{
\tau\rightarrow {a\tau+b\over c\tau+d}.}

In \GukovJK, Gukov and Witten made a proposal of how $S$-duality acts on the parameters $(\alpha_l,\beta_l,\gamma_l,\eta_l)$ labeling a half-BPS surface operator. The proposed action is  given by  \GukovJK:
\eqn\actions{\eqalign{
(\beta_l,\gamma_l)&\rightarrow |c\tau+d| \; (\beta_l,\gamma_l)\cr 
(\alpha_l,\eta_l)&\rightarrow (\alpha_l,\eta_l){\cal M}^{-1}.}}
With the aid of this proposal,  it was shown in \GukovJK\ that the gauge theory approach to the geometric Langlands program pioneered in \KapustinPK\  naturally extends to the geometric Langlands program with tame ramification. 

\medskip
\medskip
\noindent
{\it Symmetries of half-BPS Surface Operators in ${\cal N}=4$ SYM}
\medskip
\medskip


We now describe the unbroken symmetries of the half-BPS surface operators \op. 
These symmetries play an important role in determining the gravitational dual description of these operators, which we provide in the next section.

 In the absence of any insertions, ${\cal N}=4$ SYM is invariant under the $PSU(2,2|4)$ symmetry group. If we consider the surface 
 $\Sigma=R^{1,1}\subset R^{1,3}$, then $\Sigma$  breaks the $SO(2,4)$ conformal group to a subgroup. A surface operator \op\ supported on this surface inserts into the gauge theory a static probe string.
 This surface is manifestly invariant under rotations and translations in $\Sigma$ and   scale transformations. It is also invariant under the action  of inversion $I: x^\mu\rightarrow x^\mu/x^2$  and consequently\foot{We recall that  a special conformal transformation  $K_\mu$ is generated by $IP_\mu I$, where $P_\mu$ is the translation generator and $I$ is an inversion.} invariant under special conformal transformations in $\Sigma$. 
 Therefore, 
 the  symmetries left unbroken by $\Sigma=R^{1,1}$ generate an $SO(2,2)\times SO(2)_{23}$ subgroup of the $SO(2,4)$ conformal group, where $SO(2)_{23}$ rotates the plane transverse to $\Sigma$ in $R^{1,3}$.  In Euclidean signature, the surface $\Sigma=$S$^2$ preserves an $SO(1,3)\times SO(2)_{23}$ subgroup of the  Euclidean conformal group. This surface  can be obtained from the surface $\Sigma=R^{2}\in R^4$ by the action of a  broken special conformal  generator and can also be used to construct a half-BPS surface operator \op\ in ${\cal N}=4$ SYM.

Since the symmetry of a surface operator with $\Sigma=R^{1,1}$ is
$SO(2,2)\times SO(2)_{23}$ one can study 
\lref\KapustinPY{
  A.~Kapustin,
  ``Wilson-'t Hooft operators in four-dimensional gauge theories and
  Phys.\ Rev.\  D {\bf 74}, 025005 (2006)
  [arXiv:hep-th/0501015].
}
 such an operator either by
considering the gauge theory in $R^{1,3}$ or in AdS$_3\times$S$^1$, which can be obtained from $R^{1,3}$ by a conformal transformation.
Studying the gauge theory in AdS$_3\times$S$^1$ has the advantage of making the symmetries of the surface operator manifest,
\lref\GomisPG{
  J.~Gomis, J.~Gomis and K.~Kamimura,
  ``Non-relativistic superstrings: A new soluble sector of AdS(5) x S**5,''
  JHEP {\bf 0512}, 024 (2005)
  [arXiv:hep-th/0507036].
  CITATION = JHEPA,0512,024;
}
 as the conformal symmetries left  unbroken
by the surface  act by isometries on AdS$_3\times$S$^1$.
 Surface operators in $R^{1,3}$ are described by a   codimension two singularity while surface operators in AdS$_{3}\times$S$^1$ are described by a boundary condition on the boundary of AdS$_3$. A surface operator with $\Sigma=R^{1,1}$ corresponds to a boundary condition on AdS$_3$ in Poincare coordinates while a surface operator on $\Sigma=$S$^2$ corresponds to a boundary condition on global Euclidean  AdS$_3$.

 The singularity in the classical fields produced by \op\ in
 \holo\pole\ is also invariant under  $SO(2,2)$. The   ${\cal N}=4$ scalar field   $\Phi$   carries charge  under an $SO(2)_R$ subgroup of the $SO(6)$ R-symmetry and is therefore $SO(4)$ invariant. The surface operator \op\ is therefore invariant under  $SO(2,2)\times SO(2)_a\times SO(4)$, where $SO(2)_a$ is generated by the anti-diagonal product\foot{Since $SO(2)_a$ leaves $\Phi\cdot z$ in \pole\ invariant.} of $SO(2)_{23}\times SO(2)_R$. 

${\cal N}=4$ SYM has sixteen Poincare supersymmetries and sixteen conformal supersymmetries, generated by ten dimensional Majorana-Weyl spinors  $\epsilon_1$ and  $\epsilon_2$ of opposite chirality. As shown in the Appendix $A$, the surface operator ${\cal O}_\Sigma$ for $\Sigma=R^{1,1}$ preserves half of the Poincare and half of the conformal supesymmetries\foot{For $\Sigma=$S$^2$, the operator is also half-BPS, but it preserves a linear combination of Poincare and special conformal supersymmetries.} and is therefore half-BPS. 

With the aid of these symmetries we  study in the next section the gravitational description of half-BPS surface operators in ${\cal N}=4$ SYM.

\medskip
\medskip
\noindent
{\it Scaling Weight of half-BPS Surface Operators in ${\cal N}=4$ SYM}
\medskip
\medskip

Conformal symmetry constraints the form of the OPE of the energy-energy tensor $T_{mn}$ with the operators in the theory.
 For a surface operator \op\ supported on $\Sigma=R^{1,1}$, $SO(2,2)\times SO(2)_{23}$  invariance completely fixes the OPE of $T_{mn}$ with \op:
\vskip+0.5pt
 \eqn\oper{
{ <T_{\mu\nu}(x)\ope >\over <\ope  >}=h{\eta_{\mu\nu}\over r^4};\qquad 
 { <T_{ij}(x)\ope >\over <\ope >}={h\over r^4}\left[{4n_in_j-3\delta_{ij}}\right];\qquad { <T_{\mu i}(x)\ope >=0}.}
 Here $x^{m}=(x^\mu,x^i)$, where $x^\mu$ are coordinates along $\Sigma$ and $n^i=x^i/r$, and $r$ is the radial coordinate in the $R^2$ transverse to $R^{1,1}$. $h$ is the scaling weight of \op, which generalizes \KapustinPY\ the notion of conformal dimension of local conformal fields to surface operators. 
  
In order to calculate the scaling dimension of a half-BPS surface operator \op\  in ${\cal N}=4$ SYM we evaluate the classical field configuration \holo\angle\pole\ characterizing  a half-BPS surface operator on the classical energy-momentum tensor of ${\cal N}=4$ SYM:
\eqn\emten{\eqalign{
T_{mn}=&{2\over g^{2}}\Tr[D_{m}\phi D_{n}\phi-{1\over2}g_{mn}(D\phi)^2-{1\over 6}(D_{m}D_{n}
-g_{mn}D^2)\phi^2]\cr
&+{2\over g^{2}}\Tr[-F_{ml}F_{nl}+{1\over4}g_{mn}F_{lp}F_{lp}].
}}
A straightforward computation\foot{Contact terms depending on $\alpha,\beta,\gamma$ and proportional to the derivative  of the two-dimensional $\delta$-function    appear when evaluating the on-shelll energy-momentum tensor. It would be interesting to understand the physical content of these contact terms.}
leads  to:
\eqn\dimen{
h=-{2\over 3g^2}{\sum_{l=1}^MN_l(\beta_l^2+\gamma_l^2)}\,=-{1\over 6}\;{\hbox{Im}\tau\over \pi}\sum_{i=1}^MN_i(\beta_i^2+\gamma_i^2).}

 The action of an $SL(2,Z)$ transformation \fracton\  on the coupling constant of ${\cal N}=4$ SYM  implies that:
 \eqn\imt{
 \hbox{Im} \tau\rightarrow {\hbox{Im}\tau\over |c\tau+d|^2}.} 
 Combining this with the action  \actions\  of Montonen-Olive duality on the parameters of the surface operator, we find that the scaling weight \dimen\ of a half-BPS surface operator \op\ is invariant under $S$-duality:
 \eqn\invv{
 h\rightarrow h.}
 In this respect half-BPS surface operators behave like the half-BPS local operators of ${\cal N}=4$ SYM, whose conformal dimension  is invariant under $SL(2,Z)$, and  unlike   the half-BPS Wilson-'t Hooft operators  whose scaling weight is not $S$-duality invariant \KapustinPY.

\newsec{Bubbling Surface Operators}

In this section we find  the dual gravitational description of  the half-BPS surface operators \op\ described in the previous section. The bulk description  is given in terms of asymptotically \ads\  and singularity free   solutions of the Type IIB supergravity equations of motion.  The
data from which the solution is uniquely determined  encodes the corresponding data about the surface operator \op.  

The strategy to obtain these solutions is to make an ansatz for  Type IIB supergravity   which is invariant under all the symmetries preserved by the half-BPS surface operators \op. As discussed in the previous section, 
the bosonic symmetries preserved by a half-BPS surface operator \op\ are $SO(2,2)\times SO(4)\times SO(2)_a$. Therefore the most general  ten dimensional metric invariant under these symmetries  can be constructed by fibering  AdS$_3\times$S$^3\times$S$^1$ over a three manifold $X$, where the symmetries act by isometries   on the fiber. 
The constraints imposed by unbroken supersymmetry on the ansatz  are  obtained by demanding that  the ansatz for the supergravity background possesses a sixteen component Killing spinor, which means that the background  solves 
  the Killing spinor equations of Type IIB supergravity.
 A solution of  the Killing spinor equations and the Bianchi identity for the five-form field strength guarantee that the full set of equations of Type IIB supergravity are satisfied and that a half-BPS solution has been obtained. 

The problem of solving the Killing spinor equations of Type IIB supergravity   with an $SO(2,2)\times SO(4)\times SO(2)_a$ symmetry      can be obtained by analytic continuation of the equations studied by LLM \LinNB\LinNH , which found the supergravity solutions  describing  the  half-BPS local operators of ${\cal N}=4$ SYM, which have an $SO(4)\times SO(4)\times R$ symmetry. The equations determining the metric and five-form flux can be   read from \LinNB\LinNH, in which the analytic continuation that we  need to construct the gravitational description of half-BPS surface operators \op\  was considered.

The ten dimensional metric and five-form flux  is completely determined in terms of 
   data that needs to be specified on the three manifold $X$ in the ten dimensional space. An asymptotically \ads\ metric is uniquely determined in terms of a function   $z(x_1,x_2,y)$, where $(x_1,x_2,y)\equiv(\vec{x},y)$ are coordinates in $X$. The ten dimensional metric in the Einstein frame is given by\foot{The ``analytic" continuation from the bubbling geometries dual to the half-BPS local operators  is given by 
$z\rightarrow z$,
$t\rightarrow \chi$, $y \rightarrow  -iy$, ${\vec x}\rightarrow i{\vec x}$, $d
\Omega_3\rightarrow -ds^2_{AdS_3}$ \LinNB\LinNH.}
 \eqn\metric{
ds^2=y\sqrt{2z+1\over 2z-1}ds^2_{AdS_3}+y\sqrt{2z-1\over 2z+1}d\Omega_3+{2y\over \sqrt{4z^2-1}}(d\chi+V)^2+{\sqrt{4z^2-1}\over 2y}(dy^2+dx_idx_i),}
where $ds^2_X=dy^2+dx_idx_i$ with $y\geq 0$ and $V$ is a one-form in $X$ satisfying $dV=1/y *_{X}dz$. AdS$_3$ in Poincare coordinate corresponds to a surface operator on $\Sigma=R^{1,1}$ while AdS$_3$ in global Euclidean  coordinate corresponds to a surface operator on $\Sigma=$S$^2$.
 The $U(1)_a$ symmetry acts  by shifts on $\chi$ while $SO(2,2)$ and $SO(4)$ act by isometries on the coordinates of AdS$_3$ and S$^3$ respectively.  

A non-trivial solution to the equations of motion is obtained by specifying a configuration of $M$ point-like particles in $X$. The data from which the solution is determined is the ``charge" $Q_l$ of the particles together with their positions  $(\vec{x}_l,y_l)$ in $X$ (see Figure 1). Given a ``charge" distribution, the function $z(x_1,x_2,y)$ solves the following  differential equation:
\eqn\laplace{
\partial_i\partial_i z(x_1,x_2,y)+y\partial_y\left({\partial_y z(x_1,x_2,y)\over y}\right)=\sum_{l=1}^MQ_l\delta(y-y_l)\delta^{(2)}(\vec{x}-\vec{x}_l).}

Introducing a ``charge"   at the point  $(\vec{x}_l,y_l)$ in $X$ has the effect of 
shrinking\foot{Near $y=y_l$ the form of the relevant part of the metric is that of the Taub-NUT space.  Fixing the value of the ``charge" at $y=y_l$ by imposing regularity of the metric coincides with the usual regularity constraint on the periodicity of the circle in Taub-NUT space.}
 the   S$^1$ with coordinate $\chi$ in    \metric\ to zero size at that point.
  In order for this to occur in a smooth fashion  the magnitude of the ``charge" has to be fixed  \LinNB\LinNH\ so that  $Q_l=2\pi y_l $. 
 Therefore, the independent data characterizing the metric and five-form of the solution is the position of the $M$ ``charges", given by $(\vec{x}_l,y_l)$.

In summary,  a   smooth  half-BPS $SO(2,2)\times SO(4)\times SO(2)_a$  invariant asymptotically \ads\ metric  \metric\ solving the Type IIB supergravity equations of motion is found by solving \laplace\ subject to the boundary condition    $z(x_1,x_2,0)=1/2$  \LinNB\LinNH, so that the S$^3$ in \metric\ shrinks in a smooth way at $y=0$. The function $z(x_1,x_2,y)$ is   given by 
\eqn\afin{
z(x_1,x_2,y)={1\over 2}+\sum_{l=1}^M z_l(x_1,x_2,y),
}
where
\eqn\expansio{
z_l(x_1,x_2,y)={(\vec{x}-\vec{x}_l)^2+y^2+y_l ^2
\over 2
\sqrt{((\vec{x}-\vec{x}_l)^2+y^2+y_l ^2 )^2-4y_l ^2y^2}}-{1\over 2},}
and $V$ can be computed from $z(x_1,x_2,y)$ from $dV=1/y *_{X}dz$. Both the  metric and five-form field strength are determined by an integer $M$ and by   $(\vec{x}_l,y_l)$  for $l=1,\ldots, M$.  

\medskip
\medskip
\noindent
{\it Topology of Bubbling Solutions And Two-form Holonomies}
\medskip
\medskip

The asymptotically \ads\ solutions constructed from \afin\expansio\ are topologically quite rich. In particular, a solution with $M$ point ``charges" has $M$ topologically non-trivial S$^5$'s. We can associate to each point $P_l\in X$ a corresponding five-sphere S$^5_l$. S$^5_l$ can be constructed by fibering the S$^1\times$S$^3$ in the geometry \metric\ over a  straight line between the point $(\vec{x}_l,0)$ and the point $(\vec{x}_l,y_l)$ in $X$. The topology of 
this manifold is indeed an S$^5$, as an S$^5$ can be represented\foot{This can be seen explicitly by writing $d\Omega_5=\cos^2\theta d\Omega_3+d\theta^2+\sin^2\theta d\phi^2$.}
 by an S$^1\times$S$^3$ fibration over an interval where the S$^1$ and S$^3$ shrink to zero size at opposite ends of the interval, which is what happens in our geometry where the S$^3$ shrinks at $(\vec{x}_l,0)$  while the S$^1$ shrinks at the other endpoint  $(\vec{x}_l,y_l)$.

\ifig\nontrivialfive{A topologically non-trivial S$^5$ can be constructed by fibering S$^1\times$S$^3$ over an interval connecting the $y=0$ plane and the location of the ``charge" at the point $P_l\in X$ with $(\vec{x}_l,y_l)$ coordinates.}
{\epsfxsize1.5in\epsfbox{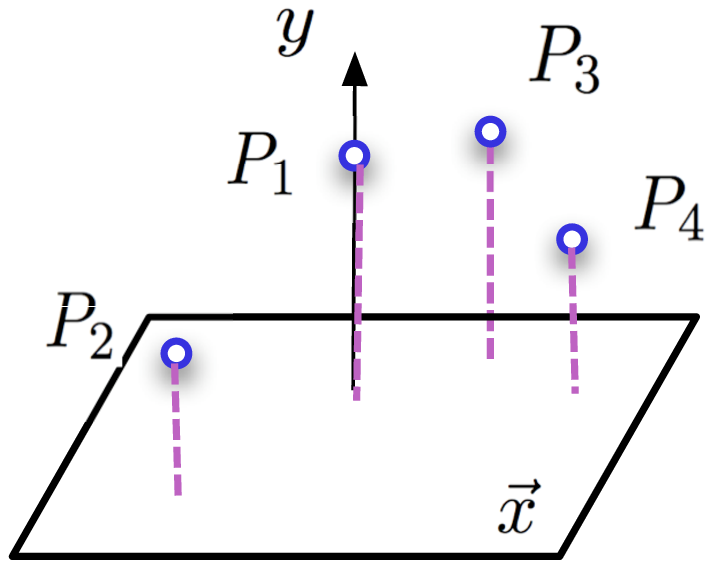}}

Following \LinNB\LinNH\ we can now integrate the five-form flux over the topologically non-trivial S$^5$'s (see Appendix $B$):
\eqn\flux{
{1\over \pi^3}\int_{S^5_l}F_5=y_l^2.}
Since flux has to be quantized, the position in the $y$-axis of the   $l$-th particle in $X$ is also quantized
\eqn\quant{
y_l^2=4\pi N_l l_p^4\qquad N_l\in Z,}
where $l_p$ is the ten dimensional Planck length. For an asymptotically \ads\ geometry with radius of curvature $R^4=4\pi Nl_p^4$, which is dual to  ${\cal N}=4$ $U(N)$ SYM, we have that the total amount of five-form flux must be $N$:
\eqn\partiti{
N=\sum_{l=1}^MN_l.}

The asymptotically \ads\ solutions constructed from \afin\expansio\ also contain non-trivial surfaces. In particular, a solution with $M$ point ``charges" has $M$ non-trivial disks $D_l$. Just as in the case of the S$^5$'s, we can associate to each point $P_l\in X$ a disk $D_l$.

Inspection of the asymptotic form of the metric \metric\ given in \afin\expansio\ reveals that   the metric is conformal to AdS$_3\times$S$^1$. This  geometry on the boundary of \ads, which is where the dual ${\cal N}=4$ $U(N)$ SYM lives, is the natural background geometry on which to study conformally invariant surface operators in ${\cal N}=4$ SYM. As explained in section $2$, an $SO(2,2)\times SO(2)_{23}$  invariant surface operator can be defined by specifying   a codimension two singularity in $R^{1,3}$ or by specifying appropriate boundary conditions for the classical fields in the gauge theory at  the boundary of AdS$_3\times$S$^1$. In the latter formulation, the worldsheet of the surface operator $\Sigma$ is the boundary of AdS$_3$.

Therefore, in the boundary of \ads\ we have a non-contractible S$^1$. 
If we fiber the S$^1$ parametrized by $\chi$ in \metric\ over a straight line connecting a point $(\vec{x}_l,y_l)$ in $X$ -- where the S$^1$ shrinks to zero size -- to a point in $X$ corresponding to the boundary of \ads\  -- given by $\vec{x},y\rightarrow \infty$ -- we obtain  a surface $D_l$. This surface is topologically a disk\foot{Such disks also appear in the
study of the high temperature regime of ${\cal N}=4$ SYM, where the bulk  geometry
\lref\WittenZW{
  E.~Witten,
  ``Anti-de Sitter space, thermal phase transition, and confinement in  gauge
  Adv.\ Theor.\ Math.\ Phys.\  {\bf 2}, 505 (1998)
  [arXiv:hep-th/9803131].
} \WittenZW\ is the AdS Schwarzschild black hole, which also has a non-contractible S$^1$ in the boundary which is contractible in the full geometry.} and there are $M$ of them for a ``charge" distribution of $M$ particles in $X$.
\ifig\nontrivialdisks{A disk $D$ can be constructed by fibering S$^1$ over an interval connecting the ``charge" at the point $P_l\in X$ with $(\vec{x}_l,y_l)$ coordinates and the boundary of \ads.}
{\epsfxsize1.5in\epsfbox{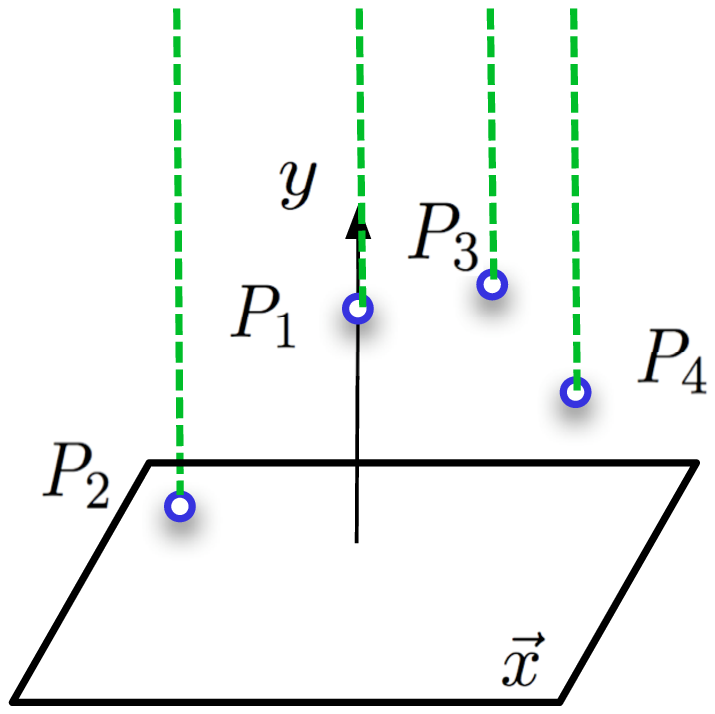}}
Due to the existence of the disks $D_i$,  the supergravity solution given by the metric and five-form flux alone is not unique. Type IIB supergravity has a two-form gauge field from the NS-NS sector and
another one from the RR sector. In order to fully specify a solution of   Type IIB supergravity in the bubbling geometry \metric\ we must complement the metric and the five-form with the integral of the two-forms around the disks\foot{The overall signs in the identification
are fixed by demanding consistent action of $S$-duality of ${\cal N}=4$ SYM with that of Type IIB supergravity.}
\eqn\holoa{
\alpha_l=-\int_{D_l} {B_{NS}\over 2\pi}\qquad \qquad \eta_l=\int_{D_l} {B_{R}\over 2\pi}\qquad l=1,\ldots,M,}
where we have used notation conducive to the later comparison with the parameters characterizing a half-BPS surface operator \op. Since both $B_{NS}$ and $B_R$ are invariant under large gauge transformations, the parameters $(\alpha_l,\eta_l)$ take values on a circle of unit radius.

Apart from the $M$ disks $D_l$, the bubbling geometry constructed from \afin\expansio\  also has   topologically non-trivial S$^2$'s. 
One can construct an S$^2$ by fibering the S$^1$ in \metric\ over a straight line connecting the points $P_l$ and $P_m$ in $X$. Since the    
S$^1$ shrinks to zero size in a smooth manner  at the endpoints  we obtain an S$^2$. Therefore, to every pair of  ``charges" in $X$, characterized by different points $P_l$ and $P_m$ in $X$, we can construct a corresponding S$^2$, which we label by S$^2_{l,m}$. The integral of $B_{NS}$ and $B_R$ over S$^2_{l,m}$ do not give rise   to new parameters, since [S$^2_{l,m}]=[D_l]-[D_m]$ in homology, and the periods can be determined from \holoa.
\ifig\nontrivialsphere{An S$^2$  can be constructed by fibering S$^1$ over an interval connecting the ``charge" at the point $P_l\in X$ with a different ``charge" at point $P_m\in X$.}
{\epsfxsize1.5in\epsfbox{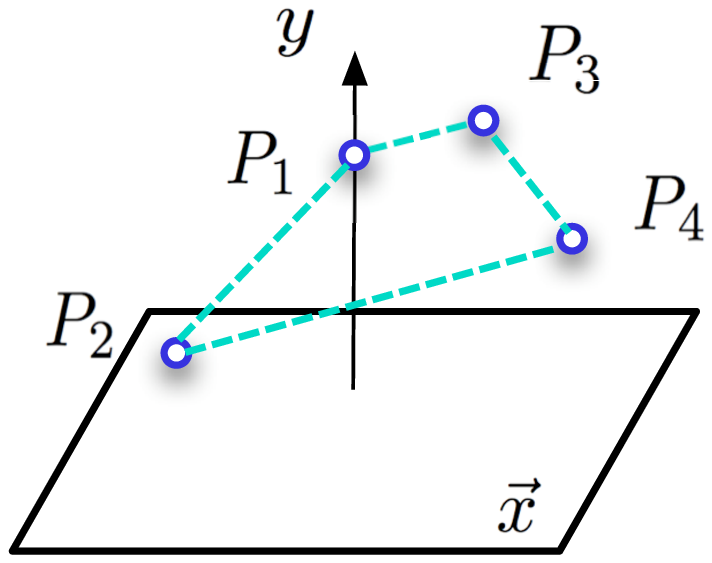}}

\medskip
\medskip
\noindent
{\it Bubbling Geometries as Surface Operators}
\medskip
\medskip

As we discussed in section $2$, a surface operator \op\  is characterized by an unbroken gauge group $L\in U(N)$ along together with $4M$ $L$-invariant parameters $(\alpha_l,\beta_l,\gamma_l,\eta_l)$. On the other hand, the Type IIB supergravity solutions we have described depend on the positions $(\vec{x}_l,y_l)$ of $M$ ``charged" particles in $X$ and the two-form holonomies:
\eqn\holob{
 \int_{D_l} {B_{NS}\over 2\pi}\qquad \qquad  \int_{D_l} {B_{R}\over 2\pi}.}
We now establish an explicit dictionary between the parameters in gauge theory and the parameters in supergravity.

For illustration  purposes,  it is convenient to start by considering the half-BPS surface operator \op\ with the largest Levi group $L$, which is $L=U(N)$ for $G=U(N)$. $U(N)$ invariance requires that the singularity in the fields produced by \op\ take values in the center of $U(N)$. Therefore, the gauge field and scalar field produced by \op\ is given by
\eqn\largest{\eqalign{
A&=\alpha_01_Nd\theta\cr
\Phi&={1\over \sqrt{2}z}(\beta_0+i\gamma_0)1_N,}}
where $1_N$ is the identity matrix. We can also turn a two-dimensional $\theta$-angle \insertion\ for the overall $U(1)$, so that:
\eqn\anglemax{
\eta=\eta_0 1_N.} 
 
We now identify this operator with the supergravity solution obtained by having a single point ``charge" source in  $X$ (see Figure 1b). If we let the position of the ``charge"   be $(\vec{x}_0,y_0)$ then 
\eqn\singlecharge{\eqalign{
z(x_1,x_2,y)&={(\vec{x}-\vec{x}_0)^2+y^2+y_0 ^2
\over 2
\sqrt{((\vec{x}-\vec{x}_0)^2+y^2+y_0 ^2 )^2-4y_0 ^2y^2}}\cr
V_I&=-\epsilon_{IJ}  
{(x^J-x^J_{0})
((\vec{x}-\vec{x}_0)^2+y^2-y_0 ^2 )
\over
2 (\vec{x}-\vec{x}_0)^2 \sqrt{((\vec{x}-\vec{x}_0)^2+y^2+y_0 ^2 )^2-4y_0
^2y^2}
},}}
where $V=V_Idx^I$.
The metric \metric\ obtained using \singlecharge\ is the metric of \ads. This can be seen by the following change of variables \LinNH
\eqn\rela{\eqalign{
x^1-x^1_0&+i(x^2-x^2_0)=r e^{i(\psi+\phi)}\cr
r&=y_0\sinh u\sin\theta\cr
y&=y_0 \cosh u \cos\theta\cr
\chi&={1\over 2}(\psi-\phi),}}
which yields the \ads\ metric with AdS$_5$ foliated by AdS$_3\times$S$^1$ slices:
\eqn\metricads{
ds^2=y_0\left[ (\cosh^2u ds^2_{AdS_3}+du^2+\sinh^2u d\psi^2)+(\cos^2\theta d\Omega_3+d\theta^2+\sin^2\theta d\phi^2)\right].}
We note that the $U(1)_a$ symmetry of the metric \metric\ --  which acts by shifts on $\chi$ -- identifies  via \rela\ an $SO(2)_R$ subgroup of the the $SO(6)$ symmetry of the S$^5$, acting by shifts on $\phi$, with an $SO(2)_{23}$ subgroup of the $SO(2,4)$ isometry group of AdS$_5$, acting by opposite shifts on $\psi$. This is precisely the same combination of generators  discussed in section $2$ that is preserved by a half-BPS surface  operator \op\ in ${\cal N}=4$ SYM.

 The radius of curvature of \ads\ in \metricads\ is given by $R^4=y_0^2$. Therefore using that $R^4=4\pi N l_p^4$, where $N$ is the rank of the ${\cal N}=4$ YM theory, we have that 
 \eqn\quanti{
 N={y_0^2\over 4\pi l_p^4},}
 and the position of the ``charge" in $y$ gets identified with the rank of the unbroken gauge group and is therefore quantized.

The residue of the pole in $\Phi$ \largest\ gets identified with the position of the ``charge" in the $\vec{x}$-plane. It follows from \metric\ that the coordinates $\vec{x}$ and $y$ have dimensions of length$^2$. Therefore, we identify the residue of the pole of $\Phi$ with the position of the ``charge" in the $\vec{x}$-plane in $X$ via:
 \eqn\mapeo{
 (\beta_0,\gamma_0)={{\vec x}_0\over 2\pi l_s^2}.}
 Unlike the position in $y$, the position in ${\vec x}$ is not quantized.
 
The remaining parameters of the surface operator \op\ with $U(N)$ Levi group -- given by  $(\alpha_0,\eta_0)$  -- get identified with the holonomy of the two-forms of Type IIB supergravity over $D$\eqn\holoc{
\alpha_0=-\int_{D} {B_{NS}\over 2\pi}\qquad \qquad \eta_0=\int_{D} {B_{R}\over 2\pi},}
where  $D$ is  the disk ending on the \ads\ boundary on the S$^1$. This  identification properly accounts for the correct periodicity of these parameters, which take values on a circle of unit radius.

The path integral which defines   a half-BPS surface operator \op\ when $L=U(N)$ is never singular as the gauge symmetry cannot be further enhanced   by changing the parameters $(\alpha_0,\beta_0,\gamma_0,\eta_0)$ of the surface operator. 
Correspondingly, the dual supergravity solution with one ``charge"  also never acquires a singularity by changing the parameters of the solution.

Let's now consider the most general half-BPS surface operator \op. First we need to characterize the operator by its Levi group, which for a  $U(N)$ gauge group takes the form $\prod_{l=1}^MU(N_l)$ with $N=\sum_{l=1}^MN_l=N$. The operator then depends on $4M$ $L$-invariant parameters   $(\alpha_l,\beta_l,\gamma_l,\eta_l)$ for $l=1.\ldots , M$ up to the action of $S_M$, which acts by permuting the parameters. 

The corresponding supergravity solution associated to such an operator is given by the metric \metric. The number of unbroken gauge group factors -- given by the integer M -- corresponds to the number of point  ``charges'' in \laplace. For $M>1$, the metric that follows from \afin\expansio\ is \ads\ only  asymptotically and   not globally. 

The rank of the various gauge group factors in the Levi group  $\prod_{l=1}^MU(N_l)$ -- given by the integers $N_l$ -- correspond to the position of the ``charges'' along $y\in X$, given by the coordinates $y_l$. The precise identification follows from \flux\quant:
\eqn\chargeiden{
N_l={y_l^2\over 4\pi l_p^4}\qquad l=1,\ldots,M.}
$N_l$ also corresponds to the amount of five-form flux over  S$^5_l$, the S$^5$ associated with the $l$-th point charge:
\eqn\chargeiden{
N_l={1\over 4\pi^4l_p^4} \int_{S^5_l}F_5\qquad l=1,\ldots,M.}
This identification quantizes the $y$ coordinate in $X$ into $l_p$ size bits. Thus length is quantized as opposed to area, which is  what happens for the geometry dual to the half-BPS local operators \LinNB\LinNH,  where it can be interpreted as the quantization of phase space in the boundary gauge theory.

A half-BPS surface operator \op\ develops a pole for the scalar field $\Phi$ \pole. The pole is characterized by its residue, which is given  by $2M$ real parameters $(\beta_l,\gamma_l)$. These parameters are identified with the position of the $M$ ``charges" in the ${\vec x}$-plane in $X$ via:
\eqn\mapeoa{
 (\beta_l,\gamma_l)={{\vec x}_l\over 2\pi l_s^2}.}
All these parameters take values on the real line.

The remaining parameters characterizing a half-BPS surface operator \op\ are the periodic variables $(\alpha_l,\eta_l)$, which determine the holonomy produced by \op\ and the corresponding two-dimensional $\theta$-angles. These parameters get identified with the holonomy of the two-forms of Type IIB supergravity over the $M$ non-trivial disks $D_l$ that the geometry generates in the presence of $M$ ``charges" in $X$:
\eqn\holoc{\eqalign{
\alpha_l&=-\int_{D_l} {B_{NS}\over 2\pi}\cr
\eta_l&=\int_{D_l} {B_{R}\over 2\pi}.}}
The identification respects the periodicity of $(\alpha_l,\eta_l)$, which   in supergravity  arises from   the invariance of $B_{NS}$ and $B_R$ under large gauge transformations\foot{Since the   gauge invariant variables  are  $e^{{i}\int_{D} {B\over 2\pi}}$.}.

We have given a complete dictionary between all the parameters that a half-BPS surface operator in ${\cal N}=4$ SYM depends on and all the parameters in the corresponding bubbling geometry. We note that a surface operator  \op\ depends on a set of parameters up to the action of the permutation group $S_M$ on the parameters, which is part of the   $U(N)$ gauge symmetry. The corresponding statement in  
 supergravity   is that the solution dual to a surface operator is invariant under the action of $S_M$, which acts by permuting the ``charges" in $X$.

The supergravity  solution is regular as long as the ``charges" do not collide. A singularity arises 
whenever two point ``charges" in $X$ coincide (see Figure 1):
\eqn\singssugra{
(\vec{x}_l,y_l)\rightarrow (\vec{x}_m,y_m)\qquad \hbox{for}\ l\neq m.}
Whenever this occurs, there is a reduction in the number of independent disks since (see Figure 3):
\eqn\singssugra{
D_l\rightarrow D_m\qquad \hbox{for}\ l\neq m,}
and therefore
\eqn\anglescoincide{\eqalign{
\int_{D_l} {B_{NS}\over 2\pi}&\rightarrow \int_{D_m} {B_{NS}\over 2\pi}\cr
\int_{D_l} {B_{R}\over 2\pi}&\rightarrow \int_{D_m} {B_{R}\over 2\pi}\qquad \hbox{for}\ l\neq m.}}
In this limit of parameter space the non-trivial S$^2$ connecting the points   $P_l$ and $P_m$ in $X$ shrinks to zero size as [S$^2_{l,m}]=[D_l]-[D_m]\rightarrow 0$, and the geometry becomes singular.

By using the dictionary developed in this paper, such a singular geometry corresponds to a limit when two    of each of the set of  parameters 
$(\alpha_l,\beta_l,\gamma_l,\eta_l)$ defining a half-BPS surface operator \op\ become equal:
\eqn\sings{
\alpha_l\rightarrow\alpha_m, \beta_l\rightarrow\beta_m, \gamma_l\rightarrow\gamma_m, \eta_l\rightarrow\eta_m\qquad \hbox{for}\ l\neq m.}
In this limit the unbroken gauge group preserved  by the surface operator \op\ is enhanced to $L'$ from the original Levi group $\prod_{l=1}^MU(N_l)$, 
 where $L\subset L'$. As explained in section $2$ the path integral from which \op\ is defined becomes singular.

 In summary, we have found the description of all half-BPS surface operators \op\ in ${\cal N}=4$ SYM in terms of solutions of Type IIB supergravity. The asymptotically \ads\ solutions    are regular 
and  when they  develop a singularity then the corresponding operator also  becomes singular.

\medskip
\medskip
\noindent
{\it $S$-Duality of Surface Operators from Type IIB Supergravity}
\medskip
\medskip

The group of dualities of ${\cal N}=4$ SYM acts non-trivially  \GukovJK\ on surface operators \op\ (see discussion in section $2$). For $G=U(N)$ the duality group is $SL(2,Z)$ and its proposed  action on the parameters on which \op\  depends on is 
\GukovJK:
\eqn\actions{\eqalign{
(\beta_l,\gamma_l)&\rightarrow |c\tau+d| \; (\beta_l,\gamma_l)\cr 
(\alpha_l,\eta_l)&\rightarrow (\alpha_l,\eta_l){\cal M}^{-1},}}
where ${\cal M}$ is an $SL(2,Z)$ matrix
\eqn\matriu{\pmatrix{a&b\cr
c&d}.}

We now reproduce\foot{If we apply the same idea to the LLM geometries dual to half-BPS local operators in \LinNB, we conclude that the  half-BPS local  operators are invariant under $S$-duality.} this transformation law by studying the action of the $SL(2,Z)$ subgroup of the $SL(2,R)$ classical symmetry of Type IIB supergravity, which is in fact the appropriate symmetry group of Type IIB string theory. For that we need to analyze the action of $S$-duality on our bubbling geometries.

  $SL(2,Z)$ acts on the complex scalar $\tau=C_0+ie^{-\phi}$ of Type IIB supergravity in the familiar  fashion
\eqn\accio{
  \tau\rightarrow {a\tau+b\over c\tau+d},}
  where as usual  $\tau$ gets identified with the complexified coupling constant of ${\cal N}=4$ SYM  \coupling.
   $SL(2,Z)$ also rotates the two-form gauge fields\foot{See e.g.
  \lref\Polchinski{
J.~Polchinski, ``String Theory", Cambridge University Press, Chapter $12$ in Volume 2.}
\lref\BBS{
K.~Becker, M.~Becker and J.~Schwarz, "String Theory and M-theory", 
Cambridge University Press, Chapter $8$.}  
\Polchinski\BBS.} of Type IIB supergravity
 \eqn\rotacion{
 \pmatrix{B_{NS}\cr
  B_R}\rightarrow 
  \pmatrix{d&c\cr
b&a}\pmatrix{B_{NS}\cr
  B_R},}
  while leaving the metric in the Einstein frame and the five-form flux invariant.
 
Given that the metric in \metric\ is in the Einstein frame, $SL(2,Z)$ acts trivially on the coordinates $({\vec x},y)$. Nevertheless, since  
\eqn\coupll{
l_s=g_s^{-1/4}l_p \qquad \hbox{with}\ g_s=e^\phi}
the string scale transforms under $SL(2,Z)$ as follows:
\eqn\transformaciosclar{
l^2_s\rightarrow  {l_s^2\over |c\tau+d|}.}
 Therefore, under $S$-duality:
 \eqn\positio{
{{\vec x}_l\over 2\pi l_s^2}\rightarrow |c\tau+d|\;{{\vec x}_l\over 2\pi  l_s^2}.}
  Given our dictionary in  \mapeoa,  we find that the surface operator parameters  $(\beta_l,\gamma_l)$ transform as in \actions, agreeing with the proposal in  \GukovJK.
 
 The identification of the rest of the parameters is \holoc:
 \eqn\holoca{\eqalign{
\alpha_l&=-\int_{D_l} {B_{NS}\over 2\pi}\cr
\eta_l&=\int_{D_l} {B_{R}\over 2\pi}.}}
Using the action   of $SL(2,Z)$ on the two-forms \rotacion\ and the identification \holoca, it follows from a straightforward manipulation that the surface operator paramaters $(\alpha_l,\eta_l)$ transform as in \actions, agreeing with the proposal in \GukovJK.

\bigbreak\bigskip\bigskip\centerline{{\bf Acknowledgements}}\nobreak
\medskip\medskip
We would like to thank Xiao Liu for very useful discussions.
Research at Perimeter Institute for Theoretical Physics is supported in
part by the Government of Canada through NSERC and by the Province of
Ontario through MRI. We also acknowledge further  support from an NSERC Discovery Grant. SM acknowledges support from JSPS Research Fellowships for Young Scientists.

\vfill\eject
\appendix{A}{Supersymmetry of Surface Operator  in ${\cal N}$=4 SYM}

In this Appendix  we study the Poincare and conformal supersymmetries  preserved by  a surface operator  in ${\cal N}$=4 SYM supported on $R^{1,1}$. These symmetries are generated by ten dimensional Majorana-Weyl spinors  $\epsilon_1$ and  $\epsilon_2$ of opposite chirality.
We determine the  supersymmetries  left unbroken by a surface operator  
by studying the supersymmetry variation of the gaugino in the 
presence of the surface operator singularity in \holo\pole.

The metric is given by:
\noindent
\eqn\minkowskico{
ds^2=-(dx^0)^2+(dx^1)^2+(dx^2)^2+(dx^3)^2=-(dx^0)^2+(dx^1)^2+2dzd\bar{z} .}
where $z={1\over \sqrt{2}}(x^2+ix^3)=|z|e^{i\theta}$, while the singularity in the fields is 
\eqn\compsca{\eqalign{
\Phi ^{\omega}&=\Phi _{\bar{\omega}} ={1\over\sqrt{2}}(\Phi^{8}+i\Phi^{9})={\beta+i\gamma \over \sqrt{2} z}\cr
A&=\alpha d\theta, F=dA=2\pi \alpha \delta_{D},}}
where  $[\alpha,\beta]=[\alpha,\gamma]=[\beta,\gamma]=0$
 and $\delta _D=d(d\theta)$
is a two form delta function. 
The relevant   $\Gamma$-matrices are 
\eqn\somedefs{\eqalign{
&\Gamma_{z}={1\over\sqrt{2}}(\Gamma_{2}-i\Gamma_{3})\cr
&\Gamma_{\bar{z}}={1\over\sqrt{2}}(\Gamma_{2}+i\Gamma_{3})\cr
&\{\Gamma_{z},\Gamma_{\bar{z}}\}=2.
}}
 
A Poincare supersymmetry transformation is  given by 
\eqnn\susygnof
$$\eqalignno{ 
\delta \lambda&= ({1\over 2}F_{\mu\nu}\Gamma ^{\mu\nu}+\nabla_{\mu}\Phi_i\Gamma^{\mu i}+{i\over 2}[\Phi_{i},\Phi_{j}]\Gamma^{ij}) \epsilon_{1} &\susygnof\cr
}
$$
where $\mu$ runs from 0 to 3 and $i$ runs from 4 to 9, while a superconformal supersymmetry transformation is given by
\eqnn\sucongnof
$$\eqalignno{ 
\delta \lambda&=[ ({1\over 2}F_{\mu\nu}\Gamma ^{\mu\nu}+\nabla_{\mu}\Phi_i\Gamma^{\mu i}+{i\over 2}[\Phi_{i},\Phi_{j}]\Gamma^{ij}) x^{\sigma}\Gamma _{\sigma}-2\Phi_{i}\Gamma^{i}] \epsilon_{2}
&\sucongnof\cr
}
$$
From \susygnof, it follows that the unbroken Poincare supersymmetries are given by:
\eqn\gaugesusy{
\Gamma ^{\bar{\omega} z}\epsilon_{1}=0 \lra \Gamma_{2389}\epsilon_{1}=-\epsilon_{1}.
}
The unbroken  superconformal supersymmetries are given by:
\eqn\confsusy{ 
\left[ \left(-{\beta+i\gamma \over z^2}\Gamma^{z \bar{\omega}}-{\beta-i\gamma \over \bar{z}^2}\Gamma^{\bar{z} \omega}\right)
 (x_0\Gamma ^{0}+x_1\Gamma ^{1}+z\Gamma ^{\bar{z}}+\bar{z}\Gamma ^{z})
-2\Phi_{\omega}\Gamma^{\omega}-2\Phi_{\bar{\omega}}\Gamma^{\bar{\omega}}\right] \epsilon_{2}=0.
}
From the terms proportional to $x^0$ and $x^1$,
we find that the unbroken superconformal supersymmetries are given by:
\eqn\suconsusy{\eqalign{
\Gamma^{z\bar{\omega}}\epsilon_{2}=0~~\lra~~\Gamma_{2389}\epsilon_{2}=-\epsilon_{2}.}}
The rest of  the conditions
\eqn\propz{
[\Gamma ^{\bar{z}}\Gamma^{\bar{\omega}}\Gamma^{z}] \epsilon_{2}=0,
}
are automatically  satisfied  once  \suconsusy\ is imposed.

We conclude that the singularity \holo\pole\ is half-BPS and that the preserved supersymmetry is generated by $\epsilon_1$ and $\epsilon_2$ subject o the constraints 
$\Gamma_{2389}\epsilon_{1}=-\epsilon_{1}$ and $\Gamma_{2389}\epsilon_{2}=-\epsilon_{2}$.

By acting with a broken special conformal transformation on $\Sigma=R^2\subset  R^4$  to get a surface operator supported on $\Sigma=$S$^2$,  
one can show following 
\lref\BianchiGZ{
  M.~Bianchi, M.~B.~Green and S.~Kovacs,
  ``Instanton corrections to circular Wilson loops in N = 4 supersymmetric
  JHEP {\bf 0204}, 040 (2002)
  [arXiv:hep-th/0202003].
}
\lref\GomisBD{
  J.~Gomis and H.~Ooguri,
  ``Non-relativistic closed string theory,''
  J.\ Math.\ Phys.\  {\bf 42}, 3127 (2001)
  [arXiv:hep-th/0009181].
}
\BianchiGZ\
that such an operator also preserves half of the thirty-two supersymmetries, but are now generated by a linear combination of the Poincare and special conformal supersymmetries.

\appendix{B}{Five form flux}

In this Appendix, we calculate \flux\
explicitly to evaluate the flux over a non-trivial S$^5$.

The five-form flux is \LinNB\LinNH
\eqn\fiveform{\eqalign{
F_5=&{1\over 4}\{d[y^2{2z+1 \over 2z-1}(d\chi+V)]-y^3\ast _3d({z+{1\over 2}\over y^2}) \}\wedge dVol _{AdS_3}\cr
&-{1\over 4}\{d[y^2{2z-1 \over 2z+1}(d\chi+V)]-y^3\ast _3d({z-{1\over 2}\over y^2}) \}\wedge d \Omega _3
}}

The five-cycle S$^5_l$ in the bubbling geometry is spanned by coordinates $\Omega _3, \chi$ and $y$.
Then the integration is:
\eqn\fiveforminteg{\eqalign{
{1\over 4\pi^4l_p^4}\int_{S^5_l}F_5 =
-{1\over 16\pi^4l_p^4} \int d[y^2{2z-1 \over 2z+1}d\chi]\wedge d \Omega _3
={y^2 _l\over 4\pi l_p^4}=N.}}

\listrefs

\end{document}

\GomisPG

\GomisCE

\GomisBD